\documentstyle[12pt]{article}
\def\pw#1{^{#1}}

\newcommand\mysection{\setcounter{equation}{0}\section}
\renewcommand{\theequation}{\thesection.\arabic{equation}}
\newcounter{hran}
\renewcommand{\thehran}{\thesection.\arabic{hran}}

\def\bmini{\setcounter{hran}{\value{equation}}
\refstepcounter{hran}\setcounter{equation}{0}
\renewcommand{\theequation}{\thehran\alph{equation}}\begin{eqnarray}}

\def\bminiG#1{\setcounter{hran}{\value{equation}}
\refstepcounter{hran}\setcounter{equation}{-1}
\renewcommand{\theequation}{\thehran\alph{equation}}
\refstepcounter{equation}\label{#1}\begin{eqnarray}}

\def\emini{\end{eqnarray}\relax\setcounter{equation}{\value{hran}}
\renewcommand{\theequation}{\thesection.\arabic{equation}}}

\def\cO#1{{\cal{O}}\!\left(#1\right)}
\def\refup#1{~$\pw{\mbox{\small #1}}$}
\def\la{\mathrel{\mathpalette\fun <}}
\def\ga{\mathrel{\mathpalette\fun >}}
\def\fun#1#2{\lower3.6pt\vbox{\baselineskip0pt\lineskip.9pt
  \ialign{$\mathsurround=0pt#1\hfil##\hfil$\crcr#2\crcr\sim\crcr}}}
\def\others{{\it et al.}}
\def\ie{\hbox{\it i.e.}{}}      
\def\eg{\hbox{\it e.g.}{}}      \def\cf{\hbox{\it cf.}{ }}
\def\eV{{\rm e\kern-0.12em V}}            
  
\def\half{{\textstyle {1\over2}}}

\def \al {\relax\ifmmode{\alpha}\else{$\alpha${ }}\fi}
    \def\Re{\mathop{\rm Re}}
\def\abs#1{\left| #1\right|}
\def\lrang#1{\left\langle #1 \right\rangle}

\def\ben{\begin{enumerate}}  \def\een{\end{enumerate}}
\def\bit{\begin{itemize}}    \def\eit{\end{itemize}}
\def\beq{\begin{equation}}   \def\eeq{\end{equation}}
\def\beeq{\begin{eqnarray}}  \def\eeeq{\end{eqnarray}}

\def\kp{\relax\ifmmode{k_\perp}\else{$k_\perp${ }}\fi}
\def\kps{\relax\ifmmode{k_\perp\pw2}\else{$k_\perp\pw2${ }}\fi}
\def \as{\relax\ifmmode\alpha_s\else{$\alpha_s${ }}\fi}

\newskip\humongous \humongous=0pt plus 1000pt minus 1000pt
\def\caja{\mathsurround=0pt}
\def\eqalign#1{\,\vcenter{\openup1\jot
\caja   \ialign{\strut \hfil$\displaystyle{##}$&$
\displaystyle{{}##}$\hfil\crcr#1\crcr}}\,}

\newif\ifdtup

\def\eqal2#1{\,\vcenter{\openup1\jot
\caja   \ialign{\strut \hfil$\displaystyle{##}$&\hfil$
\displaystyle{{}##}$\hfil &$
\displaystyle{{}##}$\hfil\crcr#1\crcr}}\,}

  \def\refup#1{~$\pw{\cite{#1}}$}
  \def\refupd#1#2{~$\pw{\citd{#1}{#2}}$}
  
  \def\refupm#1#2{~$\pw{[\ref{#1}-\ref{#2}]}$}

 \def\cite#1{[\ref{#1}]}
 \def\citd#1#2{[\ref{#1},\ref{#2}]}

\def\vkp{{\vec{k}_{\perp}}}

\def\vkps#1{\vec{k}_{\perp #1}}
\def\beql#1{\beq\label{#1}}
\def\e#1{\mathop { \mbox{e}\pw{#1}} }

\def\np#1#2#3{{\em Nucl.~Phys.}~\underline{B#1} (19#3) #2}

\def\pr#1#2#3{{\em Phys.~Rev.}~\underline{#1} (19#3) #2}

\def\zp#1#2#3{{\em Z.~Phys.}~\underline{C#1} (19#3) #2}

\begin{document}

\thispagestyle{empty}
\setcounter{page}{0}
\vbox to 1 truecm {}

\begin{flushright}
November 1994 \\[0.1cm]
LU TP 94-21 \\
LPTHE-Orsay 94-98 \\
BI-TP 94-57
\end{flushright}

\vfill
\def\cen{\centerline}

\cen{{\bf\large INDUCED GLUON RADIATION IN A  QCD MEDIUM }}
\renewcommand{\thefootnote}{\fnsymbol{footnote}}
\vskip 1.5 truecm
\centerline
{\bf R.~Baier~$\pw1$, Yu.~L.~Dokshitzer~$\pw{2,3}$, S.~Peign$\bf
{\acute e}$~$\pw3$ and D.~Schiff~$\pw3$}
\vskip 10 pt
\centerline{{\it $\pw{1 }$Fakult\"at f\"ur Physik,
Universit\"at Bielefeld, D-33501 Bielefeld, Germany}}
\centerline{{\it $\pw{2 }$ Department of Theoretical Physics,
University of Lund, S-22362 Lund, Sweden\footnote[2]
{Permanent address: Petersburg Nuclear Physics Institute, Gatchina,
 188350 St. Petersburg, Russia } }}
\centerline{{\it $\pw{3 }$
LPTHE\footnote[3] { Laboratoire associ\'e du
Centre National de la Recherche Scientifique}
, Universit\'e  Paris-Sud, B\^atiment 211, F-91405 Orsay, France}}

\renewcommand{\thefootnote}{\arabic{footnote}}
\vskip 2 cm
\cen{\bf Abstract}
\vskip 5pt
\noindent
{ Soft gluon radiation induced by multiple scattering of a fast quark
or gluon propagating through  QCD matter is discussed.
After revisiting the Landau-Pomeranchuk-Migdal effect in QED
we show that large formation times of bremsstrahlung quanta
determine the QCD radiation intensity (in analogy to QED)
and derive the gluon energy spectrum.
Coherent suppression takes place as compared to the Bethe-Heitler situation
of independent emissions.
As a result the energy loss of fast partons in a QCD medium
depends on the incident energy $E$ similarly to QED,
$-dE/dz \propto \sqrt{E}$. }

\vfill \eject

\mysection{Introduction}
The radiative energy loss encountered by a charged fast particle which
undergoes successive
scatterings has been considered long ago
in the framework of QED by Landau, Pomeranchuk and Migdal\refupm{LP}{TMAS}.
It is worth trying to reformulate the QED derivation with the aim of
a generalization to QCD. Especially, this is of importance for understanding
the energy loss mechanisms of quarks and gluons propagating through dense
matter, such as a quark-gluon plasma\refupd{BjG}{GW}.

To this end we adopt the model elaborated by Gyulassy and Wang\refup{GW}
who have recently attempted the study of radiation induced
by multiple scatterings
of a colour charge (quark) in a QCD medium.
This model disregards the collisional energy loss by introducing
static scattering centres described by the Debye screened Coulomb potential
resembling one gluon exchange.
The assumption that the mean free path $\lambda$
of the projectile is much larger than the screening radius,
$\lambda \gg \mu\pw{-1}$,
allows one to treat successive scatterings as {\em independent}
and consider the induced radiation of {\em soft}\/ gluons (photons) with
energies $\omega \ll E$.

The soft approximation essentially simplifies the derivation and leads to
an eikonal picture of classical propagation of a relativistic particle
with energy $E \gg \mu$ which receives independent elastic kicks.

The QED emission amplitude that corresponds to a single scattering
may be conveniently written in terms of a two-dimensional ``angle''
$\vec{u}={\vkp}/\omega$
with $\vkp$ the transverse momentum of the photon with respect to the direction
of the fast (massless) charge.
This may be generalized to multiple scattering and the part of the full
emission amplitude attached to
the scattering centre  \#$i$
reads
\beql{1.1}
 \vec{J}_i = \frac{\vec{u_i}}{u_i\pw2} - \frac{\vec{u}_{i-1}}{u_{i-1}\pw2}\>;
\qquad  \vec{u}_{i} = \vec{u}_{i-1} - \frac{\vec{q}_{\perp i}}{E} \>,
\eeq
where $i\!-\!1$ and $i$ mark the directions of incoming and outgoing charges
correspondingly,
and $\vec{q}_{\perp i}$ stands for the momentum transfer.

An important role is played by the interference between the amplitudes
(\ref{1.1}).
Thus  the relative eikonal phase
of the two amplitudes due to centres \#$i$ and \#$j$ reads
in the relativistic kinematics
\beql{1.2}
 k\pw\mu (x_i-x_j)_\mu \approx \sum_{m=i}\pw{j-1} \left. \left(z_m-
z_{m+1}\right)
\right/ \tau_m(k)\>; \qquad \tau_m(k) = \frac{2\omega}{(\vkps{m})\pw2}
= \frac{2}{\omega u_m\pw2}\>,
\eeq
with $z_m$ the longitudinal coordinate position of the $m\pw{\mbox{\tiny th}}$
centre.
$\tau(k)$ is usually referred to as a {\em formation time}\/ of radiation.

For {\em small}\/ formation times the phase is large and the centres act
as independent sources of radiation.
This is the Bethe-Heitler (BH) limit in which the radiation
per unit length is maximal and proportional to the single scattering spectrum,
\beql{BH}
\omega\frac{dI}{d\omega\, dz}
= \frac1{\lambda} \left(\omega\frac{dI}{d\omega}\right)_{(1)} ;
\qquad
\left(\omega\frac{dI}{d\omega}\right)_{(1)} \propto {\alpha} \>,
\eeq
with $\alpha$ the fine structure constant.
To the contrary, in the case of {\em large}\/ formation times the phase
vanishes and the amplitudes (\ref{1.1}) with $i\le m \le j $
add up coherently which results in a {\em single}\/ scattering amplitude
\beql{1.4}
 \sum_{m=i}\pw{j} \vec{J}_m =
\frac{\vec{u_j}}{u_j\pw2} - \frac{\vec{u}_{i-1}}{u_{i-1}\pw2}\>.
\eeq
The radiation density gets suppressed by $(j\!-\!i)$.
This coherent suppression ultimately
leads to the so called factorization limit.

Whether radiation follows the BH or factorization pattern,
depends on $\omega$ and properties of the medium.
In the Migdal approach to the problem of electron propagation in matter,
a Fokker-Planck type equation has
been used in which the scattering properties of the medium are
characterized by a single parameter
\beql{qdef}
 q = \left. \lrang{\Theta_s\pw2} \right/ \lambda\>; \qquad
\lrang{\Theta_s\pw2}= C\, \mu\pw2/E\pw2 \>,
\eeq
with $\lrang{\Theta_s\pw2}$
the mean squared scattering angle due to a single kick.
In agreement with qualitative arguments of \cite{LP},
the radiation intensity which is found\refup{M}
exhibits suppression
at small photon energies as compared to the BH regime (\ref{BH}):
\beql{1.6}
 \omega\frac{dI}{d\omega\, dz} \propto \alpha \sqrt{{q}\, {\omega} } \>.
\eeq
This result may be easily illustrated qualitatively,
as also discussed\refupd{FP}{GGS} using the notion
of  {\em coherence length}.
To do that we may estimate the characteristic phase (\ref{1.2}) as
\beql{1.7}
\phi \equiv \abs{k\pw\mu(x_i-x_j)_\mu } \approx \lambda \frac {\omega}
{2} \sum_{m=0}\pw{j-i-1} u_{i+m}\pw2 \>\approx\>
\frac{\lambda\omega}{2}\, \left\{ n u_{i}\pw2 + \frac{n(n\!-\!1)}{2}
\left(\frac{\mu}{E}\right)\pw2 \right\} ,
\eeq
with $n\equiv (j-i)$.
Here we treated the transverse movement of the charge as a random walk with a
typical momentum transfer $\lrang{q_{\perp}\pw2}=\mu\pw2$:
$u_{i+m}\pw2= u_i\pw2 + m \mu\pw2/E\pw2$.

Since the
QED radiation according to (\ref{1.1}) vanishes for emission angles larger than
the scattering angle,
it is legitimate to assume
$   u_{i}\pw2 \la  \lrang{\Theta_s\pw2} =   {\lrang{q_\perp\pw2}}/ {E\pw2}
    =  {\mu\pw2}/{E\pw2}$.
Therefore for large distances, $n\gg 1$, the last term in (\ref{1.7})
dominates\footnote{Strictly speaking, this is not true for the Rutherford
scattering (see below). In this case, however, the two terms of (\ref{1.7})
are of the same order, and the qualitative estimate (\ref{1.10}) still holds.}.

To keep the phase small, $\phi<1$, one has to restrict the maximal distance
between the scattering points that act coherently\refupd{FP}{GGS}, namely
\beql{1.9}
n \la N_{coh}(\omega) \equiv 2\sqrt{\frac{E\pw2}{\lambda \mu\pw2\omega}} \>.
\eeq
Treating, effectively, a group of centres with $(j\!-\!i)<N_{coh}$
as a single radiator, one arrives at the following estimate
for the radiation intensity (up to $\log$ factors)
\beql{1.10}
\omega\frac{dI}{d\omega\, dz}
= \frac1{\lambda} \left(\omega\frac{dI}{d\omega}\right)_{(1)} \cdot
\frac{1}{N_{coh}(\omega)} \> \propto\>  \frac{\alpha}{\lambda N_{coh}}
\>\propto\> \alpha\, \sqrt{\frac{\mu\pw2}{\lambda E\pw2}\, \omega}\>,
\eeq
which expression coincides with the Migdal result (\ref{1.6}).

However, in the pure Coulomb case,
for which the potential is not screened at {\em short}\/ distances,
such that $q$ becomes logarithmically divergent
for large incident energies ($q\!\sim\!\log E$),
the Migdal approach is not adequate.
In view of the application to QCD, it is important to describe
the energy spectrum of photons in the case of a potential
that has pure Coulomb behaviour at large momentum transfer\refup{D}.
Below we present the main steps of the derivation and the result
which slightly differs from the canonical one\refup{M} by an extra
logarithmic energy dependent enhancement factor
which is due to fluctuations with emission/scattering
angles much larger than typical $\mu/E$ (see (\ref{resdens}) below).

\mysection{Revisiting the LPM spectrum in QED}
The differential soft photon spectrum can be written as
\beql{2.1}
 \omega\frac{dI}{d\omega\, d\pw2u} =
\frac{\alpha}{\pi\pw2}
\lrang{ \abs{\sum_{i=1}\pw{N} \vec{J}_i \e{ik_\mu x_i\pw\mu} }\pw2}
= \frac{\alpha}{\pi\pw2}
\lrang{ 2\Re \sum_{i=1}\pw{N} \sum_{j=i+1}\pw{N} \vec{J}_i \vec{J}_j
\left[\,  \e{ik_\mu (x_i-x_j)\pw\mu} -1\,\right]
\>+\> \abs{\sum_{i=1}\pw{N} \vec{J}_i}\pw2  }
\eeq
The brackets $\lrang{ }$ indicate the averaging procedure discussed in
\cite{GW} and hereafter.
The differential energy distribution of photons radiated
{\em per unit length}\/ is given by
\beql{2.2}
 \omega\frac{dI}{d\omega\, dz} = \lambda\pw{-1}\, \frac{\alpha}{\pi}
\int\frac{d\pw2 U_0}{\pi} \> 2\Re \sum_{n=0}\pw\infty  \vec{J}_1 \vec{J}_{n+2}
\left[\,
\exp\left\{ i\kappa \sum_{\ell=1}\pw{n+1} U_\ell\pw2\,
\frac{z_{\ell+1}-z_\ell}{\lambda} \right\} -1 \,\right] .
\eeq
Here we have expressed the relative phases in terms of the
{\em rescaled}\/ angular variable,
\beql{Udef}
 \vec{U} \equiv \vec{u} \cdot \frac{E}{\mu}\>, \quad
\left( \vec{u}\equiv \frac{\vkp}{\omega} \right) ,
\eeq
which measures the relative photon angle in units of
a typical scattering angle $\Theta_1=\mu/E$.
We have also introduced the characteristic parameter
\beql{kapdef}
\kappa = N_{coh}\pw{-2}= m\half \lambda \mu\pw2 \frac{\omega}{E\pw2} \> \ll \>
1\>.
\eeq
In (\ref{2.2}) we have taken advantage of the facts that 1)
the last term in (\ref{2.1}), corresponding to the factorization
limit remains finite and therefore
does not contribute to the radiation {\em density}, and
2)  the internal sum becomes $i$-independent for large $N$.
The latter property allows one to suppress the index $\#i$
in (\ref{2.2}) and keep only the number of kicks $n$ between the
scattering points $1$ (former $\#i$) and $n\!+\!2$ ($\#j$).

It is worthwhile to notice that only the expression
integrated over photon angles has a well defined
$N\!\to\!\infty$ limit.
Meanwhile, the {\em angular spectrum}\/ per unit length
is meaningless since the direction of the charge inside the medium,
which enters in the definition of
$\vec{U}_0$ cannot be precisely specified.

For the averaging procedure, two steps have to be implemented in (\ref{2.2}).
First, one has to integrate over the longitudinal separation between successive
scattering centres $\Delta_\ell = z_{\ell+1}-z_\ell$ with the normalized
distribution probability
\beql{zdist}
 \prod_{\ell=1}\pw{n+1}  \frac{d\Delta_\ell}{\lambda}
 \exp\left(-\frac{\Delta_\ell}{\lambda}\right)\>.
\eeq
This results in substituting for the square brackets
in (\ref{2.2}) the expression
\beql{psiprod}
 \prod_{\ell=1}\pw{n+1} \psi(U_\ell\pw2) \>-\>1\>; \qquad
 \psi(U\pw2) = \left(1-i\kappa U\pw2\right)\pw{-1}\>.
\eeq
Secondly, an averaging over momentum transfers $\vec{q}_{\perp\ell}$
should be performed with the distribution corresponding to the screened
Coulomb potential scattering:
\beql{qpdist}
 \prod_{\ell=1}\pw{n+2} dV(\vec{q}_\ell)\>; \qquad
dV(\vec{q}_\ell) = \frac{\mu\pw2\, d\pw2q_\ell}{\pi (q_\ell\pw2 + \mu\pw2)\pw2}
\>;
\quad \int dV(\vec{q}_\ell) = 1\>.
\eeq
Corresponding integrals may be evaluated analytically in the
large angle approximation, $U\pw2\gg 1$ (that is, $u\pw2\gg (\mu/E)\pw2$),
which region {\em a posteriori}\/ proves
to give the main contribution to the energy spectrum\refup{D}.

The result may be written in the following form
\beql{nres}
 \omega\frac{dI}{d\omega dz} = \lambda\pw{-1} \frac{\alpha}{\pi}
\int_0\pw\infty
\frac{dU\pw2}{U\pw2+1} \> 2\Re \sum_{n=0}\pw\infty
\left[\, f_n(U\pw2;0) - f_n(U\pw2;\kappa) \,\right] ,
\eeq
where the $n$-kick contributions satisfy approximately
the recurrency relation
\bminiG{recrel}
 f_{n+1}(U\pw2) &=& \psi(U\pw2) \, f_n(U\pw2+1) \cdot
\left\{ 1 + \cO{U\pw{-4}} \right\} , \\
 f_0(U\pw2) &=& \frac{\psi(U\pw2)}{U\pw2(U\pw2+1)} \>; \qquad
 f_n(U\pw2) = \frac{\prod_{\ell=0}\pw{n}
\psi(U\pw2+\ell)}{(U\pw2+n)(U\pw2+n+1)}\>.
\emini
The dependence on $\kappa$ enters implicitly via the
$\psi$ factors.
The characteristic number of terms essential in (\ref{nres}) is estimated as
\beql{nest}
\lrang{n} \sim U\pw2 \>.
\eeq
Finally, one arrives\refup{D} at
\beql{Ures}
\eqalign{
\omega\frac{dI}{d\omega dz} &
= 2\lambda\pw{-1} \frac{\alpha}{\pi} \int_0\pw\infty
\frac{dU\pw2}{U\pw2(U\pw2+1)} \> \Phi(\kappa U\pw4) \>; \cr
\Phi(x) &= \int_0\pw\infty \frac{dt}{(t+1)\pw{3/2}} \,
\sin\pw2 \frac{x\,t}4 \>.
}\eeq
For large values of the argument $\kappa U\pw4 \sim
  \lrang{n}\pw2/N_{coh}\pw2 >1$
the $\Phi$ factor tends to unity which corresponds to the BH limit,
which is in accord with the qualitative estimate given in the previous section.
In the contrary, in the region $U\pw2\ll \kappa\pw{-1/2}$ the BH law gets
modified,
\beql{modif}\eqalign{
 \Phi(x) & \approx
\frac{\sqrt{\pi}}{2} \, \sqrt{x}\>, \quad (x\ll 1)\>;\cr
 \omega\frac{dI}{d\omega dz} & \approx   \frac{\alpha}{\pi}
\frac{\sqrt{\pi\kappa}}{\lambda} \int_0\pw{1/\sqrt{\kappa}}
\frac{dU\pw2}{U\pw2+1}\>.
}\eeq
As a result, the main contribution originates from a broad logarithmic region
\beql{logreg}
  1 \ll U\pw2 \ll \kappa\pw{-1/2} \>.
\eeq
The resulting radiation density reads
\beql{resdens}
 \omega\frac{dI}{d\omega dz} \approx \frac{\alpha}{\pi}
 \frac{\sqrt{\pi\kappa}}{2\lambda} \ln (\kappa\pw{-1})
= \frac{\alpha}{2\lambda}\sqrt{\frac{\lambda\mu\pw2\, \omega}{2\pi\>  E\pw2}}
\, \ln\frac{2\,E\pw2}{\lambda\mu\pw2  \omega} \>.
\eeq
The above derivation has been based on the approximation
$1\ll \lrang{n}\sim U\pw2 < N_{coh}=\kappa\pw{-1/2} $.
This implies that the photon spectrum
is suppressed  for energies below a specific value given by:
\bminiG{ests}
\label{QEDomBH}
 \kappa < 1 \>\>\Longleftrightarrow\>\>  \omega <
\omega_{BH}=\frac{2E\pw2}{\mu\pw2\lambda}
\equiv E\left(\frac{E}{E_{LPM} }\right) , \quad
E_{LPM} = \half \mu\pw2\lambda\>.
\eeeq
Photons with $\omega > \omega_{BH}$ are radiated according to the BH law.

On the other hand, we treated the medium as being large enough
in the longitudinal direction to embody $N_{coh}$ successive scatterings.
For a medium of finite size $L$ this condition limits photon energies from
below,
\beeq
\label{QEDomfact}
 \lambda N_{coh} < L  \>\>\Longleftrightarrow\>\>
  \omega > \omega_{fact}=\frac{\lambda E\pw2}{L\pw2\mu\pw2} \equiv
E \left( \frac{L_{cr}}{L} \right)\pw2  , \quad
L_{cr}= \sqrt{\frac{\lambda E}{\mu\pw2}}\>.
\emini
Photon radiation with $\omega < \omega_{fact}$ corresponds to
the factorization limit.

The LPM effect may be observed in the photon energy range
\beq
   \frac{L\pw2_{cr}}{L\pw2}  < \frac{\omega}{E} <
  \frac{E}{E_{LPM} }  \>,
\eeq
which obviously demands the size of the medium to be big enough,
$L > \max \left\{ \lambda, \sqrt{\lambda E / \mu\pw2} \right\}$.

\mysection{ LPM effect and  energy dependent energy loss in a QCD
medium}
In this section we consider gluon radiation in a QCD medium,
induced by multiple scattering of an energetic quark
due to one gluon exchange with a static centre.
An essential feature of the QCD model
is that the accompanying radiation does not vanish in the $E\!\to\!\infty$
limit as it was the case for QED scattering.
The reason is simple:
gluon emission
does not vanish in the case of forward scattering, $\Theta_s\!\to\!0$,
due to ``repainting'' of the incident quark via colour exchange.

This makes it possible to simplify the treatment and consider
the quark as moving along the $z$ axis.
In such an approximation the QED-like part of the
induced radiation gets suppressed and only
the dominant specific non-Abelian contribution survives.
The radiation amplitude for the latter is proportional to
the commutator of colour generators, $[T\pw{a} , T\pw{b}]=if_{abc}T\pw{c}$,
with $a,b$ the colour indices of the emitted gluon
and of the octet potential, correspondingly.
In the soft approximation the basic amplitude reads
\beql{qcdamp}
 \vec{J}_i = \frac{\vkp}{\kps}
-\frac{\vkp-\vec{q}_{\perp i}}{(\vkp-\vec{q}_{\perp i})\pw2} \,.
\eeq
It combines the three contributions with the same eikonal phase
$\exp(i z_i \kps/2\omega)$
adjusted to the position of the centre $\#i$,
which originate from Feynman diagrams for
the gluon emission off incoming and outgoing quark lines
as well as from the exchanged gluon line (via 3-gluon coupling).

Hereafter we depict the effective eikonal current
(\ref{qcdamp})
by a blob attached to the gluon line (as shown in Fig.\ref{fig3} below)
which convention is chosen to stress the $if_{abc}$ colour
structure of the current.

Gauge invariance can be enforced, as discussed in detail in [\ref{BDPS}].
To this end we remark here that the only way to construct
a simple self-consistent first order QCD model that embodies
the screened gluon potential
is to ascribe to the gluon field a finite mass $\mu$ which acts
as screening parameter for space-like exchange and as
particle mass for the real emission.
In the following, for the sake of simplification,
we shall be using, however, the unmodified current
(\ref{qcdamp}) which a posteriori will be justified by the fact
that the resulting expression for the induced radiation density
proves to be collinear safe and dominated by $\kps\gg\mu\pw2$.

\subsection{Gyulassy-Wang treatment of induced gluon radiation}
As a first step, we focus on quark rescattering processes in the
spirit of Gyulassy-Wang's approach\refup{GW}.

The derivation basically follows that of the QED case, with the
only essential difference coming from the fact that
the contributions associated to successive quark rescatterings
between $\#i$ and $\#j$ centres acquire colour factors
$(-(2N_cC_F)\pw{-1})\pw{j-i-1} = (-N_c\pw2+1)\pw{-(j-i-1)}$.
This leads to colour suppression as implied by the nonplanar nature
of the corresponding diagrams.

As long as it is not the angle but the transverse momentum
which is relevant for the emission amplitude
(cf. (\ref{1.1}) and (\ref{qcdamp})),
we express the eikonal phase difference (\ref{1.2}) in terms
of $(\vkps{m})\pw2=\kps$ which remains $m$-independent
within the adopted approximation ($E\!\gg\!\mu$).
After averaging over the position of scattering centres with
the weight (\ref{zdist}) where $\lambda$ now stands for the
quark mean free path, the $\psi$ factors emerge
\beql{psiqcd}
\psi(\kps) = \left(1-i\kappa  {\kps} / {\mu\pw2}
 \right)\pw{-1} \>\equiv\> \left(1-i\kappa U\pw2 \right)\pw{-1}\>.
\eeq
Here we have introduced the characteristic QCD parameter
\beql{kappaQCDdef}
  \kappa = \half {\lambda\mu\pw2}/{\omega} \>\ll\> 1\>.
\eeq
The two dimensional vector $\vec{U}$ now stands
for the gluon transverse momentum measured in units of $\mu$.

To integrate the product of the currents
$\vec{J}_i\vec{J}_j$ over transferred momenta
$\vec{q}_i, \vec{q}_j$        one uses
\beql{prop}
\int\frac{d\pw2q}{\pi} \frac{\vkp -\vec{q}}{(\vkp -\vec{q}\,)\pw2}
\, f(q\pw2) = \frac{\vkp}{\kps} \int_0\pw{\kps} dq\pw2\, f(q\pw2)\>,
\eeq
and arrives at the radiation density
\beql{rdQCD1}\eqalign{
\omega\frac{dI}{d\omega \, dz}
&= \lambda\pw{-1} \frac{N_c\as}{\pi}
 \int_0\pw\infty \frac{dU\pw2}{U\pw2(U\pw2+1)\pw2} \>
\Re \left[\, g(1) - g(\psi(U\pw2)) \, \right] \>; \cr
g(\psi) &\equiv \frac{N_c}{2C_F} \psi \sum_{n=0}\pw\infty
\left(  -\frac{\psi}{2N_cC_F} \right)\pw{n}
=  \frac{\psi}{1+(\psi-1)/N_c\pw2} \>; \quad (n=j-i-1)\,.
}\eeq
It is clear that in the large $N_c$ limit the answer
is determined by the
interference between the nearest neighbours, $j=i+1$.
Straightforward algebra leads to
\beql{rdQCD2}
\frac{\omega\, dI}{d\omega \, dz}
= \frac{N_c\as}{\pi\,\lambda}
 \int\limits_0\pw\infty \frac{U\pw2\, dU\pw2}{(U\pw2\!+\!1)\pw2}
\left[\, U\pw4\!+\!
\left( \frac{N_c\, \omega}{C_F \lambda\mu\pw2} \right)\pw2  \,\right]\pw{-1}
\! =   \frac{N_c\as}{\pi\,\lambda}  \int\limits_0\pw\infty
\frac{\mu\pw4\>d\kps }{\kps (\kps\!+\!\mu\pw2)\pw2} \, \left[\,
1 \!+\! \left( \frac{N_c}{2C_F}\, \frac{\tau}{\lambda}
\right)\pw2\,\right]\pw{-1}
\eeq
Except in the case $\lambda\mu\pw2\gg E$ where one
formally
recovers the BH
limit, (\ref{rdQCD2}) leads to
a sharply falling gluon energy distribution
$\omega dI/d\omega \propto \omega\pw{-2}
$
and, therefore, to finite energy losses
\beq
  -{dE}/{dz} \sim \mbox{const}\cdot \as\, \mu\pw2 \>.
\eeq
This qualitatively agrees, up to $\log E$ factors,
with the conclusions of~\cite{GW}.

The origin of this result is quite simple:
in the quark rescattering model the gluon radiation
with formation time $\tau$ (\cf (\ref{1.2}))
exceeding $\lambda$ is negligible,
which severely limits gluon energies:
$$
 \omega =  {1 \over 2} \tau \kps \la \lambda q_\perp\pw2 \sim \lambda
\mu\pw2\>.
$$
The fact that the quark after having emitted a gluon at point $\#i$
effectively stops interacting with the medium looks as if it
had lost its ``colour charge''.
The truth is, however, that the colour charge has not disappeared
but has been transferred to the radiated gluon.

\subsection{Radiated gluon rescattering in the large Nc limit and the
analogy with QED}
The diagrams for the product of the emission currents
$\vec{J}_i\vec{J}_j$
which are displayed in Fig.\ref{fig3} dominate in the large $N_c$ limit.

\begin{figure}
 \setlength{\unitlength}{0.5pt}
 \thicklines
\newsavebox{\macroC}
\savebox{\macroC}(0,0)[bl]{
\put(  0.00,  0.00){\circle*{5}}   }
\newsavebox{\macroD}
\savebox{\macroD}(0,0)[bl]{
\put(  0.00,  0.00){\circle*{10}}  }
\newsavebox{\maccross}
\savebox{\maccross}(0,0)[bl]{
\put(-10, 10){\line(1,-1){20}}
\put(-10,-10){\line(1, 1){20}}     }
\newsavebox{\macroA}
\savebox{\macroA}(0,0)[bl]{
\put( 67.00, 67.00){\oval( 24.00, 78.00)[r]}
\multiput(-383.00,174.00)( 56.00,  0.00){5}{\vector(1,0){ 55.00}}
\multiput(-331.00,-23.00)( 56.00,  0.00){5}{\vector(-1,0){ 55.00}}
\put(-359.26,174.00){\oval(  9.48,  9.33)[bl]}
\put(-359.26,164.67){\oval(  9.48,  9.33)[tr]}
\put(-363.79,164.67){\oval( 18.52,  9.33)[br]}
\put(-363.79,153.00){\oval( 23.12, 14.00)[tl]}
\put(-363.57,153.00){\oval( 23.55, 14.00)[bl]}
\multiput(-363.57,146.00)(  0.43,-28.00){5}{
\put(  0.00, -7.00){\oval( 28.21, 14.00)[tr]}
\put(  0.21, -7.00){\oval( 27.79, 14.00)[br]}
\put(  0.21,-21.00){\oval( 27.79, 14.00)[tl]}
\put(  0.43,-21.00){\oval( 28.21, 14.00)[bl]}  }
\put(-361.43, -1.00){\oval( 23.55, 14.00)[tr]}
\put(-361.21, -1.00){\oval( 23.12, 14.00)[br]}
\put(-361.21,-12.67){\oval( 18.52,  9.33)[tl]}
\put(-365.74,-12.67){\oval(  9.48,  9.33)[bl]}
\put(-365.74,-22.00){\oval(  9.48,  9.33)[tr]}
 \multiput(-355.00, 54.00)( 56,  0.10){4}{\usebox{\maccross}}
\thinlines
\multiput(-381.00, 54.00)( 41.30,  0.10){10}{           
\put(  0.00,  0.00){\line(1,0){ 30.98}} }
\multiput(-299.00,-23.00)( 56.00,  0.00){3}{
\put( -3.10,  0.00){\oval(  6.19,  6.10)[tr]}
\put( -3.10,  6.10){\oval(  6.19,  6.10)[bl]}
\put( -0.14,  6.10){\oval( 12.10,  6.10)[tl]}
\put( -0.14, 13.71){\oval( 15.10,  9.14)[br]}
\put( -0.29, 13.71){\oval( 15.38,  9.14)[tr]}
\multiput( -0.29, 18.29)( -0.29, 18.29){5}{
\put(  0.00,  4.57){\oval( 18.43,  9.14)[bl]}
\put( -0.14,  4.57){\oval( 18.14,  9.14)[tl]}
\put( -0.14, 13.71){\oval( 18.14,  9.14)[br]}           
\put( -0.29, 13.71){\oval( 18.43,  9.14)[tr]}
}
\put( -1.71,114.29){\oval( 15.38,  9.14)[bl]}
\put( -1.86,114.29){\oval( 15.10,  9.14)[tl]}
\put( -1.86,121.90){\oval( 12.10,  6.10)[br]}
\put(  1.10,121.90){\oval(  6.19,  6.10)[tr]}
\put(  1.10,128.00){\oval(  6.19,  6.10)[bl]}
}
\thicklines
\multiput(-334.00,146.00)( 19.67,  0.00){9}{
\put(  0.00, -19.67){\oval( 32.78, 39.33)[tr]}
\put(  9.83, -19.67){\oval( 13.11, 39.33)[b]}
\put( 19.67, -19.67){\oval( 32.78, 39.33)[tl]}   }
\put(-335.00,139.00){\oval( 25.00, 14.00)[tl]}
\put(-64.00,173.00){\vector(1,0){ 36.00}}
\put(-28.00,173.00){\vector(1,0){ 51.00}}
\put( 26.00,-23.00){\vector(-1,0){26}}
\put( 0,-23.00){\vector(-1,0){ 50.00}}

\multiput(-125.00,144.00)( 13.00,  0.00){3}{\usebox{\macroC}}
\multiput(-125.00,  4.00)( 13.00,  0.00){3}{\usebox{\macroC}}
\multiput(-72.00,147.00)( 18.75,  0.00){5}{
\put(  0.00, -18.75){\oval( 31.25, 37.50)[tr]}
\put(  9.38, -18.75){\oval( 12.50, 37.50)[b]}
\put( 18.75, -18.75){\oval( 31.25, 37.50)[tl]}     }
\multiput( 23.00,147.00)( 14.67,-13.67){3}{
\put(  0.00,-28.33){\oval( 24.44, 56.67)[tr]}
\put(  7.33,-28.33){\oval(  9.78, 29.33)[b]}
\put( 14.67,-28.33){\oval( 24.44, 29.33)[tl]}      }
\multiput(  0.00,  0.00)( 16.75,  7.00){4}{
\put(  0.00, 23.75){\oval( 27.92, 47.50)[br]}
\put(  8.38, 23.75){\oval( 11.17, 33.50)[t]}
\put( 16.75, 23.75){\oval( 27.92, 33.50)[bl]}      }
\multiput(-302.00,106.00)( 57.00,  1.00){3}{\usebox{\macroC}}
\put( -361.00, -23.00){\usebox{\macroC}}
\multiput( -301.00, -23.00)( 57.00,  0.00){3}{\usebox{\macroC}}
\put( -364.00,174.00){\usebox{\macroC}}
\put( -350.00,138.00){\usebox{\macroD}}
}

\begin{center}
\begin{picture}(500,220)
\put(395.00, 39.00){\usebox{\macroA}}
  \put(382.00, 46.00){\usebox{\macroD}}
  \put(395.00, 44.00){\oval(20,10)[bl]}

\put(378.03,212.00){\oval( 10.05, 10.05)[bl]}
\put(378.03,201.95){\oval( 10.05, 10.05)[tr]}
\put(373.00,201.95){\oval( 20.10, 10.05)[br]}
\put(373.00,189.38){\oval( 25.13, 15.08)[tl]}
\put(373.00,189.38){\oval( 25.13, 15.08)[bl]}
\multiput(373.00,181.85)(  0.00,-30.15){4}{
\put(  0.00, -7.54){\oval( 30.15, 15.08)[tr]}
\put(  0.00, -7.54){\oval( 30.15, 15.08)[br]}
\put(  0.00,-22.62){\oval( 30.15, 15.08)[tl]}
\put(  0.00,-22.62){\oval( 30.15, 15.08)[bl]}
}
\put(373.00, 53.69){\oval( 30.15, 15.08)[tr]}
\put(373.00, 53.69){\oval( 30.15, 15.08)[br]}
\put(373.00, 38.62){\oval( 25.13, 15.08)[tl]}
\put(373.00, 38.62){\oval( 25.13, 15.08)[bl]}
\put(373.00, 26.05){\oval( 20.10, 10.05)[tr]}
\put(378.03, 26.05){\oval( 10.05, 10.05)[br]}
\put(378.03, 16.00){\oval( 10.05, 10.05)[tl]}
\put(373.03,212.00){\usebox{\macroC}}
\put(373.03,16){\usebox{\macroC}}
\put(365,93){\usebox{\maccross}}
\end{picture}
\end{center}

\begin{center}
\begin{picture}(500,220)
\put(395.00, 39.00){\usebox{\macroA}}
  \put(380.00, 42.00){\usebox{\macroD}}
  \put(395.00, 44.00){\oval(24,10)[bl]}
\put(369,148.00){\usebox{\macroC}}
\put(368.38,16){\usebox{\macroC}}
\put(365,93){\usebox{\maccross}}
\put(373.38,147.00){\oval(  6.77,  6.77)[bl]}
\put(373.38,140.23){\oval(  6.77,  6.77)[tr]}
\put(370.00,140.23){\oval( 13.54,  6.77)[br]}
\put(370.00,131.77){\oval( 16.92, 10.15)[tl]}
\put(370.00,131.77){\oval( 16.92, 10.15)[bl]}
\multiput(370.00,126.69)(  0.00,-20.31){4}{
\put(  0.00, -5.08){\oval( 20.31, 10.15)[tr]}
\put(  0.00, -5.08){\oval( 20.31, 10.15)[br]}
\put(  0.00,-15.23){\oval( 20.31, 10.15)[tl]}
\put(  0.00,-15.23){\oval( 20.31, 10.15)[bl]}         }
\put(370.00, 40.38){\oval( 20.31, 10.15)[tr]}
\put(370.00, 40.38){\oval( 20.31, 10.15)[br]}
\put(370.00, 30.23){\oval( 16.92, 10.15)[tl]}
\put(370.00, 30.23){\oval( 16.92, 10.15)[bl]}
\put(370.00, 21.77){\oval( 13.54,  6.77)[tr]}
\put(373.38, 21.77){\oval(  6.77,  6.77)[br]}
\put(373.38, 15.00){\oval(  6.77,  6.77)[tl]}
\end{picture}
\end{center}

    \caption{ \label{fig3}
     Eikonal graphs with gluon rescattering dominate in the large $N_c$
     limit.
     The part of the diagram below the dashed line corresponds to the
     conjugate emission amplitude.
     Crosses mark the scattering centres.}
    \end{figure}
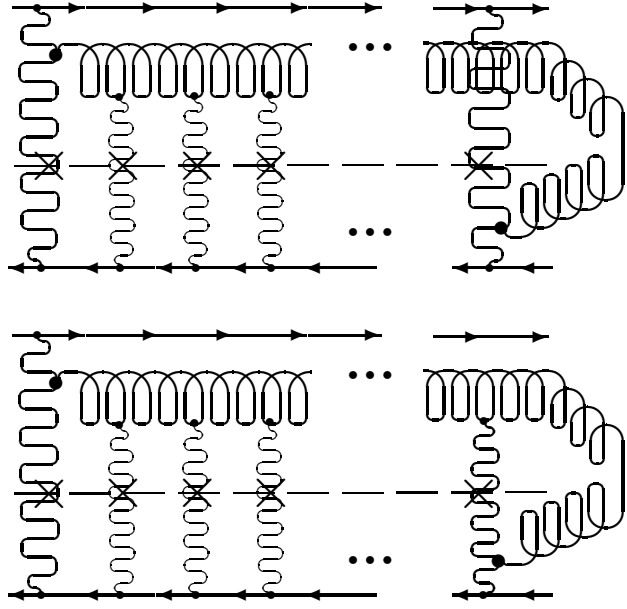

    \begin{figure}
\begin{center}
\begin{picture}(400,150)
\thicklines
\put(-50,-40){
\put(40,50){$q_i$}
\put(30,110){$k\!-\!q_i$}
\put(100,80){$k$}
\put(355.00, 95.00){\circle*{2.5}}
\multiput(357.00, 95.00)(  0.00,-16.00){3}{
\put(  0.00, -4.00){\oval( 16.00,  8.00)[tl]}
\put(  0.00, -4.00){\oval( 16.00,  8.00)[bl]}
\put(  0.00,-12.00){\oval( 16.00,  8.00)[tr]}
\put(  0.00,-12.00){\oval( 16.00,  8.00)[br]}        }
\put(357.00, 43.00){\oval( 16.00,  8.00)[tl]}
\put(357.00, 43.00){\oval( 16.00,  8.00)[bl]}

 \newsavebox{\macroE}
\savebox{\macroE}(0,0)[bl]{
\put( 23.00,  0.00){\vector(1,0){ 56.00}}
\put( 79.00,  0.00){\line(1,0){ 20.00}}
\put( 46.40,  0.00){\oval(  4.81,  4.81)[bl]}
\put( 46.40, -4.81){\oval(  4.81,  4.81)[tr]}
\put( 44.00, -4.81){\oval(  9.62,  4.81)[br]}
\put( 44.00,-10.82){\oval( 12.02,  7.21)[tl]}
\put( 44.00,-10.82){\oval( 12.02,  7.21)[bl]}
\multiput( 44.00,-14.43)(  0.00,-14.43){6}{
\put(  0.00, -3.61){\oval( 14.43,  7.21)[tr]}
\put(  0.00, -3.61){\oval( 14.43,  7.21)[br]}
\put(  0.00,-10.82){\oval( 14.43,  7.21)[tl]}
\put(  0.00,-10.82){\oval( 14.43,  7.21)[bl]}        }
\multiput( 44.00,  0.00)(  1.00,-43.00){2}{\circle*{2.5}}
\multiput( 56.00,-43.00)(  7.80,  0.20){5}{
\put(  0.00,  8.00){\oval( 13.00, 16.00)[br]}
\put(  3.90,  8.00){\oval(  5.20, 15.60)[t]}
\put(  7.80,  8.00){\oval( 13.00, 15.60)[bl]}        }
\put( 44.00,-43.00){\line(1,0){ 12.00}}
\put(  0.00,  0.00){\vector(1,0){ 23.00}}            }
\put(180.00,137.00){\usebox{\macroE}}
\put( 24.00,137.00){\usebox{\macroE}}
\put(224.00, 95.00){\circle*{6}}
\put(150,70){$\displaystyle \Longrightarrow$}
\put(290,70){$\displaystyle +$}

\put(307.00,137.00){\vector(1,0){ 42.00}}
\put(349.00,137.00){\line(1,0){ 48.00}}
\put(308,150){$z_{i\!-\!1}$}
\multiput(307.00,137.00)(  7.60, -8.40){5}{
\put(  0.00,-16.00){\oval( 12.67, 32.00)[tr]}
\put(  3.80,-16.00){\oval(  5.07, 15.20)[b]}
\put(  7.60,-16.00){\oval( 12.67, 15.20)[tl]}         }
\put(345.00, 95.00){\line(1,0){ 12.00}}

\multiput(362.00, 95.00)(  7.80,  0.20){5}{
\put(  0.00,  8.00){\oval( 13.00, 16.00)[br]}
\put(  3.90,  8.00){\oval(  5.20, 15.60)[t]}
\put(  7.80,  8.00){\oval( 13.00, 15.60)[bl]}          }
\put(355.00, 95.00){\line(1,0){ 7.00}}
\multiput(224.00,160.00)(131.00,  1.00){2}{
\multiput(  0.00,  0.00)(  0.00,-19.25){7}{
\put(  0.00,  0.00){\line(0,-1){ 14.44}}              }
\put(5,-10){$z_{i}$}           }      }

\end{picture}
\end{center}
    \caption{\label{fig2}
    Feynman 3--gluon diagram produces two eikonal graphs:
    it participates in the gluon production current
    (second term of the amplitude  (3.1))
    and gives rise to Coulomb rescattering.}
    \end{figure}
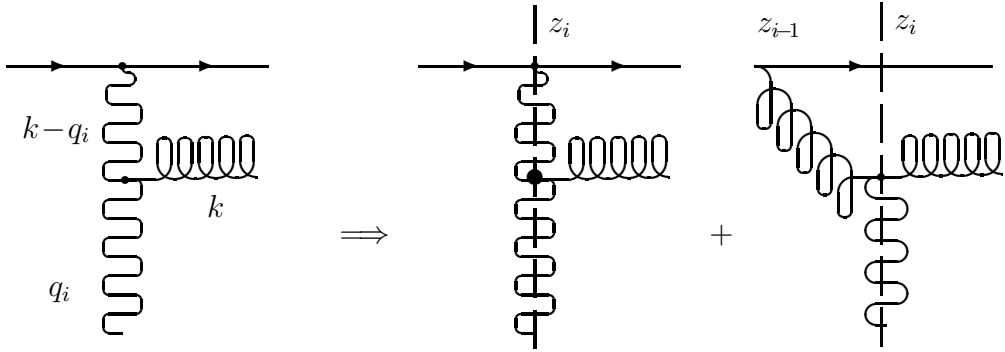

Let us consider the Feynman diagrams with
gluon self-interaction as shown in Fig.\ref{fig2}.
Integration over the position of the quark-gluon interaction vertex $t$
between the successive scattering centres, $z_{i-1}\le t\le z_i$,
gives rise to two contributions with the phase factors attached to
$z_i$ and $z_{i-1}$.
The first term in the rhs of Fig.\ref{fig2}
corresponds to an instantaneous interaction mediated by the
{\em virtual}\/ (Coulomb) gluon $(k\!-\!q_i)$ and is responsible for
the second term of the basic scattering amplitude (\ref{qcdamp}).
The second one corresponds to the production of the {\em real}\/ gluon
$(k\!-\!q_i)$ in the previous interaction point, which then rescatters
at the centre $\#i$.

The sum of these two terms is proportional to the difference
of the two phase factors.
Therefore when phases are set to zero (factorization limit),
gluon production {\em inside}\/ the medium cancels.
The only contributions which survive correspond to the gluon radiation
at the very first and the very last interaction vertices.
As a result, the subtraction term analogous to $ |{\sum \vec{J}_i}|\pw2 $
of (\ref{2.1}) remains $N$-independent and does not contribute
to the radiation {\em density}.
In the QCD case this statement is less transparent than in QED,
since the initially produced gluon is still subject to
multiple rescattering in the medium.
The latter, however, does not affect the {\em energy}\/ distribution
of radiation.

A similar argument applies to the final state interaction after
the point $\#j$ which has been neglected in Fig.\ref{fig3}:
Coulomb interaction of the quark-gluon pair with the medium
does affect the transverse momentum distribution of the
gluon but does not change its energy spectrum,
the only quantity which concerns us here.

It is worthwhile to mention that in the QCD problem the very notion of
scattering cross section (and, thus, of mean free path $\lambda$)
becomes elusive: the scattering cross section of the quark-gluon pair
depends on the transverse size of the 2--particle system and therefore
is $z$-dependent.
As we shall see below,
the radiation intensity of gluons with $\omega\!\gg\!\lambda\mu\pw2$
is due to comparatively {\em large}\/ formation times
$\tau\!\gg\!\lambda$
but not large enough for the quark and the gluon to separate in the
impact parameter space.
Such a pair as a whole acts as if it were a single quark propagating
through the medium.
Therefore the mean free path of the quark, $\lambda\propto C_F\pw{-1}$,
may be effectively used for the eikonal averaging in the spirit of
(\ref{zdist}).
Better formalized considerations will be given elsewhere\refup{BDPS}.

The emission current (\ref{qcdamp}) and the structure of the diffusion
in the variable $\vec{u}$ prove to be identical for
the QCD and QED problems.
The only but essential difference is that now $\vec{u}$ has to be related
to the {\em transverse momentum}\/ instead of the {\em angle}\/ of the gluon:
\beql{anal}
\eqalign{
\mbox{QED:}~~ &
 \vec{u}_i = \frac{\vkps{i}}{\omega},
\> \vec{U}_i= \vec{u}_i\frac{E}{\mu}; \>
 \kappa = \frac{\lambda\mu\pw2}{2} \,\frac{\omega}{E\pw2} \ll 1
\quad \Longrightarrow \quad
\cr
  \mbox{QCD:}~~ &
\vec{u}_i = \vkps{i}, \> \vec{U}_i= \frac{\vec{u}_{i}}{\mu} ;
\>
 \kappa =  \,\frac{\lambda\mu\pw2} {2\,\omega} \ll 1\, .
}\eeq
Correspondingly, the $\kappa$ parameter gets modified according to the
QCD expression (\ref{kappaQCDdef}).

Given this analogy,
the QCD derivation follows that of the previous section.
The graph of Fig.\ref{fig3}b  is twice the graph of Fig.\ref{fig3}a
due to colour factors.
One finds the gluon energy distribution (\cf (\ref{Ures}))
\beql{UresQCD}
\omega\frac{dI}{d\omega dz} =
\frac{3 N_c\as}{\pi\,\lambda}
\int_0\pw\infty \frac{dU\pw2}{U\pw2(U\pw2+1)} \> \Phi(\kappa U\pw4) \>\approx\>
\frac{3 N_c\as}{4\lambda}\sqrt{\frac{\lambda\mu\pw2}{2\pi\omega}}
\, \ln\frac{2\omega}{\lambda\mu\pw2} \>.
\eeq

\subsection{Rescattering of the incident parton and the complete spectrum}
Now we are in a position to complete the analysis of induced QCD
radiation by taking into account subdominant quark rescattering
contributions.

Each gluon scattering provides the phase factor $\psi(U\pw2)$
accompanied by the colour factor $N_c/2$
which one normalizes by the quark elastic scattering factor
 $C_F\pw{-1}\propto \lambda_q$.
Each quark scattering graph supplies the colour factor
$(-1/2N_c)\cdot C_F\pw{-1} = (C_F-N_c/2)C_F\pw{-1}$
and the same $\psi$ factor (since the gluon momentum
stays unaffected, and the change in the quark direction is negligible).
Accounting for an arbitrary number of quark rescatterings in-between
two successive gluon rescatterings results in a modified phase factor:
\bminiG{lamtil}
\label{lamtila}
\frac{N_c}{2C_F} \psi \cdot
 \sum_{m=0}\pw\infty \psi\pw{m}\left(1 - \frac{N_c}{2C_F}\right)\pw{m}
&=&
\left(1+ \frac{2C_F}{N_c}\left[\,\psi\pw{-1}-1 \,\right] \right)\pw{-1}
\equiv \tilde{\psi} \>; \\
\label{lamtilb}
\tilde{\psi}(U\pw2) = \left( 1 - i\tilde{\kappa} U\pw2 \right)\pw{-1} ;
&&   \!\!\!
\tilde{\kappa}\equiv \frac{\tilde{\lambda} \mu\pw2}{2\omega} \>, \>\>
C_F \lambda_q \equiv \frac{N_c}{2}\, \tilde{\lambda} \,. \qquad{}
\emini
It is worthwhile to notice that the modified effective mean free path
$\tilde{\lambda}$
does not depend on the nature of the colour representation
$R$ of the initial particle.
For example,
in the case of the {\em gluon}\/ substituted for the initial quark,
Coulomb rescattering of both projectile and radiated gluons supplies
the colour factor $N_c/2$,
while the elastic cross section provides the normalization
$N_c\pw{-1}\propto \lambda_g$.
Thus, since $N_c\lambda_g=C_F \lambda_q $,
the same result follows:
\beq
 \frac{N_c}{2} N_c\pw{-1}\psi \cdot \sum_{m=0}\pw\infty
\left( \frac{N_c}{2} N_c\pw{-1}\psi \right)\pw{m} \>=\> \tilde{\psi}\>.
\eeq
In general, for an arbitrary colour state $R$ one has to replace
$C_F$ in (\ref{lamtila}) by a proper Casimir operator $C_R$
which dependence then cancels against the $C_R\pw{-1}$ factor that
enters the expression for the mean free path of a particle $R$:
$C_F \lambda_q =N_c\lambda_g =C_R\lambda_R= \half N_c\tilde{\lambda}$.

Thus, (\ref{UresQCD}) holds
provided one replaces $\kappa$ by the modified
$\tilde{\kappa}$ according to (\ref{lamtilb}).
To factor out the dependence on the type of the projectile one may express
the answer in terms of the gluon mean free path
$\lambda_g=\half\tilde{\lambda}$:
\beql{QCDres1}
 \omega\frac{dI}{d\omega dz} = 3\cdot
\frac{C_R \as}{\pi\,\lambda_{g}}
\int_0\pw\infty \frac{dU\pw2}{U\pw2(U\pw2+1)} \> \Phi(\tilde{\kappa} U\pw4)
\>\approx\>
 \frac{3C_R \as}{4\, \lambda_g}
\sqrt{\frac{\lambda_g\mu\pw2}{\pi\, \omega}}
\, \left(\ln\frac{\omega}{\lambda_g\mu\pw2} \>+\> \cO{1}  \right) .
\eeq
This is our final result for the QCD induced radiation spectrum.
It has been derived with logarithmic accuracy
in the energy region
\beql{cond}
 {\omega}\left/{\lambda_g\mu\pw2} \right. = \tilde{\kappa}\pw{-1} \>\gg \> 1\>.
\eeq

It is this condition which, a posteriori, justifies the use of the quark
scattering cross section for the quark-gluon system.
Indeed, the random walk estimate for the transverse separation
between $q$ and $g$ in course of $\lrang{n}\la 1/\sqrt{\kappa}$ kicks
gives
\beql{separ}
\Delta \vec{\rho}_\perp \approx \lambda \sum_{m=1}\pw{\lrang{n}}
{\vkps{m}}/ {\omega} \>: \quad
(\Delta \vec{\rho}_\perp)\pw2
\sim \frac{\lambda\pw2\mu\pw2}{\omega\pw2} \lrang{n}\pw2
\la \kappa \, \mu\pw{-2}\> \ll\> \mu\pw{-2}\,.
\eeq
Given the separation which is much {\em smaller}\/ than the radius of the
potential, the mean free path of the $qg$ system in the medium
coincides with that of a single quark.

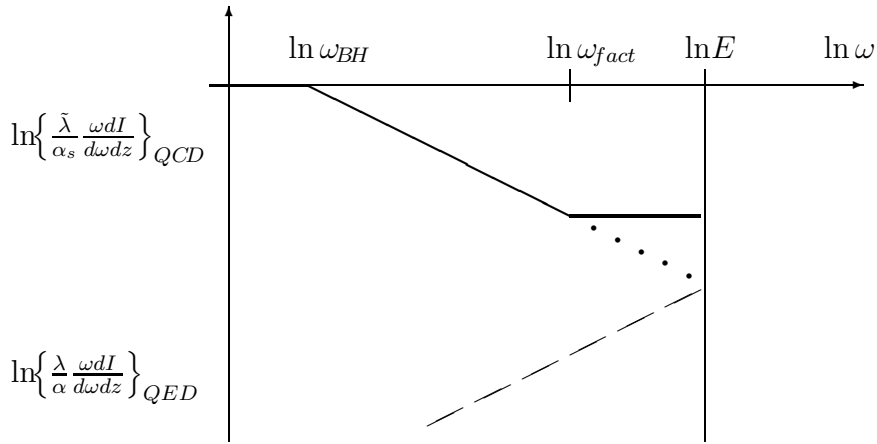
\begin{figure}
\setlength{\unitlength}{1.5pt}
\begin{center}
\begin{picture}(200,115)
\thinlines
\put(50,90){
\put(-20,-90){\vector(0,1){110}}
\put(-20,0){\vector(1,0){160}}
\multiput(30,-86)(12,6){6}{\line(2,1){9}}
\put(100,-90){\line(0,1){94}}
\put(66,-4){\line(0,1){8}}
\thicklines
\put(-25,0){\line(1,0){25}}
\put(0,0){\line(2,-1){66}}
\put(66,-33){\line(1,0){33}}
\multiput(72,-36)(6,-3){5}{\circle*{1}}
\put(95,7){$\ln\! {E} $}
\put(130,7){$\ln {\omega} $}
\put(60,7){$\ln {\omega_{\!fact}} $}
\put(-5,7){$\ln {\omega_{\!B\!H}} $}
\put(-75,-15){$\ln\!\!\left\{
\frac{\tilde{\lambda}} {\as}\frac{\omega dI}{d\omega dz}\right\}
_{Q C\!D} $}
\put(-75,-75){$\ln\!\!\left\{
\frac{\lambda}{\alpha}\frac{\omega dI}{d\omega dz}\right\}
_{Q E D}$}    }
\end{picture}
\end{center}\caption{\label{figex}
The normalized QCD radiation density (solid line) for a finite size medium
($L\!<\!L_{cr}$).
The QED LPM spectrum ($E>E_{LPM}$) is shown by the dashed line. }
\end{figure}

A comment concerning the factorization and BH regime limits of the
spectrum is in order and proceeds along the same lines as in QED
(see (\ref{ests})).
For the medium of a finite size $L\!<\!L_{cr}=\sqrt{\lambda_g E/\mu\pw2}$
the $\omega$-independent factorization limit holds for energetic gluons with
\bminiG{BHfac}
\omega \ga \omega_{fact}=E\left( {L} / {L_{cr}}\right)\pw2 , \qquad
\left( N_{coh}=\tilde{\kappa}\pw{-1/2} \ga L/\lambda_g \right).
\eeeq
The BH regime corresponds formally to small gluon energies
\footnote{
Notice that for $\kappa\!\gg\! 1$ the separation between the colour charges
according to (\ref{separ}) becomes  much {\em  larger}\/ than $1/\mu$.
In these circumstances the normalization cross section tends to the
sum of independent $q$ and $g$ contributions, $C_F \Longrightarrow
C_F \!+\! N_c \approx 3 C_F$.
The factor 3 in (\ref{QCDres1})
disappears leading to the standard BH expression. },
\beeq
 \omega \la \omega_{BH}= \lambda_g\mu\pw2\,, \qquad (\tilde{\kappa} \ga 1)\>.
\emini
The spectrum (\ref{QCDres1}) is depicted in  Fig.~\ref{figex}  together
with the QED spectrum.
When it is  integrated over $\omega$ up to $E$, it leads to energy losses
\beq
 -\frac{dE}{dz} \>\propto\> C_R\as \sqrt{E\, \frac{\mu\pw2}{\lambda_g} } \>
\ln\frac{E}{\lambda_g\mu\pw2} \qquad (\mbox{for}\>\> L>L_{cr})\, .
\eeq

\mysection{Discussion and concluding remarks}
In this letter we have considered induced soft gluon radiation
off a fast colour charge propagating through the medium
composed of static QCD Coulomb centres.

Due to the colour nature of the scattering potential,
the specific non-Abelian contributions to the gluon yield dominate
over the QED-like radiation for all gluon energies (up to $\omega \la E$).
They have been
singled out here by choosing the $E\!\to\!\infty$
limit in which the Abelian radiation vanishes.

Having established a close analogy between the angular structure of the QED
problem and the transverse momentum structure of the QCD problem
one can qualitatively obtain the gluon energy density spectrum
from the known QED result by the substitution
$\omega/E\pw2\to 1/\omega$.
The spectrum of gluon radiation is $E$-independent and,
analogously to the Landau-Pomeranchuk-Migdal effect in QED,
acquires coherent suppression
as compared to the Bethe-Heitler regime of independent emissions:
\beql{4.1}
  \omega\frac{dI}{d\omega\, dz}\propto \lambda_g\pw{-1}
\as \cdot\sqrt\frac{\lambda_g\mu\pw2}{\omega}\,
\ln\frac{\omega}{\lambda_g\mu\pw2} \>, \quad
\left(\frac{\lambda_g\mu\pw2}{\omega}\equiv \tilde{\kappa}
=N_{coh}\pw{-2} \ll 1\right).
\eeq
As a result, the radiative energy loss per unit length
amounts to
\beql{4.2}
-dE/dz \propto \as\sqrt{E \mu\pw2/\lambda_g}\ln(E/\lambda_g\mu\pw2) .
\eeq
\noindent
This is in apparent contradiction with the statements
existing in the literature on the subject.
In particular, it contradicts the recent conclusion presented in \cite{GW}
about the {\em finite}\/ density of energy losses.

As discussed above, the latter statement
is due to a sharply falling energy spectrum
$\omega  {dI}/{d\omega dz}$ $\propto \omega\pw{-2}$ which one finds
within the approach disregarding Coulomb rescattering of the radiated
gluon.
Only gluon radiation with restricted formation time $\tau\la\lambda$
survives such a treatment, while (\ref{4.1}) is dominated by
large formation times $\lambda \ll \tau \ll \lambda N_{coh}$.

It is interesting to notice, that in spite of the fact that large momentum
transfers do not essentially contribute to the scattering cross section,
$d\sigma/d q\pw2_\perp\propto q_\perp\pw{-4}$ for  $q_\perp\pw2\gg\mu\pw2$,
the answer is actually related to {\em hard}\/ scatterings.
The logarithmic enhancement factor in (\ref{4.1}) originates from
interference between two radiation amplitudes:
one is related to a ``typical'' scattering with $q\pw2_n\sim \mu\pw2$
while the other corresponds to a hard fluctuation
with a very large momentum transfer up to
$q_1\pw2 \sim \mu\pw2 N_{coh}\gg \mu\pw2$.
These amplitudes get a chance to interfere because of the random walk
in the gluon transverse momentum in a course of
$n\sim \kps/\mu\pw2 \la N_{coh}$ rescatterings
with typical $q_m\pw2\sim \mu\pw2$.
Such a fluctuation actually is not rare:  the probability
that at least one of $n$  scatterings will supply the momentum transfer
$q_\perp\pw2$
exceeding $n\mu\pw2$ amounts to $1- (1\!-\!1/n)\pw{n} \sim 1$.

Our results should be directly applicable to ``hot'' (deconfined) plasma
(with $\lambda \gg 1/\mu$) in which case the model potential
(\ref{qpdist}) with $\mu$ as a screening parameter could be taken
at its face value.
The structure of the spectrum (\ref{4.1}) is such that
one has to know, strictly speaking,
the mean free path $\lambda$ separately from the screening
mass $\mu$ at high temperature.
This is in contrast to the Migdal approach extended to QCD\refup{BDPS},
in which the rhs of (\ref{4.1}) is given by $\as\sqrt{q/\omega}$,
\ie\ the plasma properties only enter via the transport coefficient
$q\equiv \lrang{q_\perp\pw2}/\lambda$.

At the same time, one might think of applying the above consideration
to the propagation of a fast quark/gluon through ``cold'' nuclear matter.
In such a case the parameter $\mu$ has to be introduced by hand as a lower
bound for transverse momenta which one may treat perturbatively.
Given the ill-defined nature of $\mu$ it is worthwhile to notice that
the inverse size of the potential and the mean free path enter (\ref{4.1})
in the combination $\mu\pw2/\lambda_g=\mu\pw2\,\rho\sigma_g$, with
$\sigma_g$ the gluon scattering cross section.
This makes it possible to express the results
in terms of the physical density of scattering centers $\rho$ and
the dimensionless parameter
$\theta\!=\!{\sigma_g}\mu\pw2  \propto  N_c\as\pw2$
which measures the strength of gluon interaction with
the medium and should be much less sensitive to
a finite
uncertainty in~$\mu$.

\vspace{1.5 cm}
\noindent
{\bf\large Acknowledgement}

Discussions with P.~Aurenche, S.J.~Brodsky, V.N.~Gribov,
A.H.~Mueller and K.~Redlich are kindly acknowledged.
One of us (D) is grateful for hospitality extended to him at the
High Energy Physics Lab. of the University Paris-Sud, Orsay
where this study has been performed.

This research is supported in part by the EEC Programme "Human Capital
and Mobility", Network "Physics at High Energy Colliders", Contract
CHRX-CT93-0357.

\newpage
\vbox to 2 truecm {}

\def\labelenumi{[\arabic{enumi}]}
\noindent
{\bf\large References}
\ben
\item\label{LP}
  L.D.~Landau and I.Ya.~Pomeranchuk,
  {\em Dokl.~Akad.~Nauk SSSR}\/  \underline {92} (1953) 535, 735.
\item\label{M}
  A.B.~Migdal, \pr{103}{1811}{56}; and references therein.
\item\label{FP}
 E.L.~Feinberg and I.Ya.~Pomeranchuk,
  {\em Nuovo Cimento Suppl.}\/ III, (1956) 652.
\item\label{TMAS}
 See, \eg:
  M.L.~Ter-Mikaelian, High Energy Electromagnetic Processes in Condensed
  Media, John Wiley \& Sons, 1972; \\
  A.I.~Akhiezer and N.F.~Shul'ga, {\em Sov.~Phys.~Usp.}\/
  \underline {30} (1987) 197;\\
                 and references therein.
\item\label{BjG}
 J.D.~Bjorken, Fermilab report PUB--82/59--THY, 1982 (unpublished); \\
M.~Gyulassy \others, {\em Nucl.~ Phys.} \underline {A 538} (1992) 37c; \\
                 and references therein.
\item\label{GW} M.~Gyulassy and X.-N.~Wang, \np{420}{583}{94}; \\
  see also
    X.-N.~Wang, M.~Gyulassy and M.~ Pl\"umer, preprint LBL--35980, August 1994.
\item\label{GGS}
 V.M.~Galitsky and I.I.~Gurevich,
 {\em Nuovo Cimento}\/ XXXII (1964) 396; \\
A.H.~S{\o}rensen, \zp{53}{595}{92}.
\item\label{D} Yu.L.~Dokshitzer, in preparation.
\item\label{BDPS} R.~Baier, Yu.L.~Dokshitzer, S.~Peign\'e and D.~Schiff,
 in  preparation.
\een

\end{document}